# Beyond Demographics:
# Behavioural Segmentation and Spatial Analytics to Enhance Visitor Experience at The British Museum

Naomi Muggleton[1,2], Timothy Monteath[1,3], and Taha Yasseri[1,4,5*]

[1]*Alan Turing Institute for Data Science and AI, London, UK*
[2]*Warwick Business School, University of Warwick, Coventry, UK*
[3]*Centre for Interdisciplinary Methodologies, University of Warwick, Coventry, UK*
[4]*Oxford Internet Institute, University of Oxford, Oxford, UK*
[5]*Centre for Sociology of Humans and Machines, Trinity College Dublin and Technological University Dublin, Dublin, Ireland*

**Abstract**

This study explores visitor behaviour at The British Museum using data science methods applied to novel sources, including audio guide usage logs and TripAdvisor reviews. Analysing 42,000 visitor journeys and over 50,000 reviews, we identify key drivers of satisfaction, segment visitors by behavioural patterns, examine tour engagement, model spatial navigation, and investigate room popularity. Behavioural clustering uncovered four distinct visitor types: Committed Trekkers, Leisurely Explorers, Targeted Visitors, and Speedy Samplers, each characterised by different levels of engagement and movement. Tour usage analysis revealed high drop-off rates and variation in completion rates across different language groups. Spatial flow modelling revealed that accessibility and proximity, particularly aversion to stairs, shaped visitor paths more than thematic organisation. Room popularity was more strongly predicted by physical accessibility than curatorial content. We propose practical strategies for improving engagement and flow, offering a scalable framework for visitor-centred, data-informed museum planning.

# Introduction

Museums and cultural institutions play a significant role as centres of public learning, cultural exchange, and the preservation of human heritage. Their missions increasingly extend beyond traditional conservation and education roles to encompass visitor engagement, personalised experiences, and responsiveness to diverse audiences (Hooper-Greenhill, 2000). This shift reflects an increasing desire to centre visitors in institutional missions, and to support their intellectual, emotional, and social experiences within the museum context (Vergo, 1989; Ridge, 2014; Hesmondhalgh, 2008). These developments have coincided with growing pressures on institutions to demonstrate public value and accountability, particularly in the context of public funding and technological disruption (Barrett, 2011; Bennett, 2013; Falk, 2016a; Peers and Brown, 2003).

*Corresponding Author: Taha Yasseri, Centre for Sociology of Humans and Machines (SOHAM), Trinity College Dublin, 2-4 Foster Pl, D02 A3K7 Dublin, Ireland. taha.yasseri@tcd.ie.

To meet these demands, museums are adopting audience-centred strategies that leverage behavioural data to inform operational and curatorial decisions. Understanding visitor behaviour in museums has long been a subject of research across disciplines, including museum studies, human-computer interaction, tourism, cultural analytics and the humanities. Yet despite substantial existing research in visitor studies, many traditional approaches relying heavily on post-visit surveys or demographic segmentation are often limited by coarse granularity, biases associated with self-reporting, and are frequently disconnected from actual in-museum behaviour (Barrett, 2011; Falk, 2016b; Hooper-Greenhill, 2000).

Emerging technologies, including mobile applications, digital audio guides, and visitor-tracking sensors, now all offer institutions the ability to capture detailed interactions and movements in real time, offering unprecedented opportunities to understand visitor preferences, behaviours, and needs (Ballantyne & Uzzell, 2011). Recent methodological advances in museum analytics have begun leveraging digital footprints, including RFID tags, Bluetooth beacons, and mobile app telemetry, to reconstruct detailed visitor paths and interactions (Ballantyne & Uzzell, 2011). These data streams offer more profound insights into spatial movement, visitor flows, and engagement patterns (Hoare, 2025).

Yet, practical applications of these insights to inform museum strategy remain limited, and there is a clear gap in integrating visitor segmentation with spatial analytics to drive strategic decision-making directly. As the adoption of these technologies progresses, museums are increasingly moving from reductive, purely data-driven, and metric-focused evaluations to a more holistic data-informed approach, where insights from visitor data can be used to guide strategic decisions in enhancing visitor experiences and operational efficiency through interpretative analytic research (Hoare, 2025).

Our work builds on these strands by combining spatial log data (from audio guides) with unstructured textual reviews. Unlike earlier work focused on single-room dwell times or event-specific flows, we explore museum-wide navigation over longer periods, as part of groups, and connect physical behaviour with sentiment analysis and clustering. Our methodology is informed by both qualitative and quantitative approaches and aims to provide museum practitioners with both theoretical and practical tools (Ridge, 2014).

Furthermore, we position our analysis within a broader shift toward data-informed cultural strategy. Museums today are increasingly expected to justify curatorial decisions and spatial designs based on real-time audience insight and public value. This study contributes to that effort by using empirical data to recommend strategic changes that align with actual visitor behaviour.

As one of the world's most visited cultural institutions - with over six million visitors annually - the British Museum continuously seeks innovative ways to improve visitors' satisfaction, manage crowd dynamics, and optimise spatial layout through strategic insights from digital data. This study investigates the context of visitor experiences at the Museum using data science-informed methods. The Museum provided the authors with two data sources of visitor data: audio guide usage logs and TripAdvisor review texts, which they knew contained rich insights into visitor interactions with the Museum, but which they lacked the resources and technical expertise to fully analyse 'in-house'.

The audio guide data provided captured the journey of over 42,000 visitors between 2016 and 2017, roughly 10% of total users of the device during that time. These logs reveal how visitors move through the Museum, how long they stay, which objects they interact with, and whether they travel alone or in groups. The TripAdvisor data, on the other hand, provides a rich, multilingual corpus of visitor reviews, offering insight into user sentiment, preferences, and satisfaction. The use of review platforms like TripAdvisor in cultural research has grown in recent years, offering a scalable source of qualitative feedback. Studies have analysed sentiment trends, lexical diversity, and cultural framing across global heritage sites (Nuccio and Bertacchini, 2023). While these platforms reflect only a subset of the visitor population, they provide valuable insights into subjective experiences, linguistic diversity, and emotional resonance. In combination, these two data sources allowed us a unique angle for exploring visitors' interactions with the Museum.

We therefore use these complementary datasets to answer four interrelated questions: (1) How can museum visitors be meaningfully segmented based on observed behavioural patterns rather than demographic characteristics? (2) What spatial and physical factors significantly influence visitor flow and room visitation within the museum? (3) Which spatial or behavioural barriers limit visitor exploration, and how can these be strategically mitigated?; and (4) How can behavioural and spatial insights directly inform museum strategy to enhance visitor experience, satisfaction, and operational efficiency?

Through these questions, our study makes several contributions. First, it introduces a novel application of clustering and movement analysis to real-world museum behaviour, aligning with recent interest in dynamic segmentation techniques for audience insight.

Second, it presents the case that spatial constraints and behavioural tendencies, rather than curatorial themes or content, are dominant factors in determining visitor flow and satisfaction. A finding that contributes to the growing body of literature questioning static interpretive models in museum layout design (Hoare, 2025; Ridge, 2014). Third, we propose a suite of practical, data-informed recommendations for museum planning, tour design, and accessibility enhancements. These align with current calls to leverage audience data for inclusive and adaptive experiences across the cultural sector, of which museums are an integral part.

Furthermore, we position our analysis within a broader push toward data-driven cultural strategy. Museums today are being asked to justify programming, optimise layouts, and expand accessibility, not only based only on curatorial intent but through demonstrable visitor-centred outcomes. This study contributes to that effort by using empirical data to recommend strategic changes that align with actual visitor behaviour.

## Methods

### Data Sources

This study leverages three complementary datasets collected at The British Museum and shared with the authors under terms and conditions. The full dataset contains the following.

#### *Audio Guide Usage Logs*

We analysed anonymous logs from handheld audio guides provided to visitors between July 2016 and October 2017, covering 42,261 unique, anonymised visitor journeys. Each log recorded timestamps, interactions with museum objects, session durations, and language settings. This dataset represents roughly 10% of audio-guide users in this period.

*TripAdvisor Reviews*

We analysed 52,514 visitor reviews of The British Museum from TripAdvisor, spanning multiple years and 38 languages. Reviews were collected by the Museum and supplied as part of the collaboration. Each review contained numerical ratings (15 stars), textual comments (review title and main text), and optional metadata such as user nationality, visit date, and trip type (e.g., solo, couple, business).

*Google Analytics Summary Data*

Aggregate statistics from the British Museum's Google Analytics account were used to cross-reference seasonal visitor trends and page views per object, helping validate findings from the audio guide and review datasets.

These datasets are complementary: the audio logs reveal fine-grained behavioural traces within the building, while TripAdvisor reviews capture reflective, emotional, and qualitative aspects of the visitor experience. All data was used in accordance with data protection and ethical research and data protection standards – both by The British Museum prior to delivering the data to us and during our research conducted at the Alan Turing Institute. This included the anonymisation of personal identifiers, secure data storage, and adherence to user data usage agreements. In practice, this means that individuals cannot be identified from the data, their participation was voluntary and subject to data protection legislation.

## Data Preprocessing and Cleaning

Before analysis, all datasets underwent significant preprocessing and filtering to ensure quality and consistency.

*Audio Guide Data Cleaning*

We removed sessions with fewer than three object interactions or less than five minutes in duration to exclude test runs, accidental activations, or technical errors. Sessions were split into individual trips, accounting for cases where devices were shared between people or reused without reset. Anomalous patterns, such as jumping between distant locations in short intervals, were flagged and excluded.

Each trip was associated with metadata including start time, total number of stops, language used, estimated group size (inferred from device sharing), and tour completion if applicable. All object IDs were mapped to room locations and cultural themes using an internal object-room database provided by the Museum.

*TripAdvisor Review Parsing*

Reviews were translated into English using machine translation when not originally written in English using the R package translateR (Lucas & Tingley, 2014). Ratings were normalised to a five-star scale. Trip types (e.g., solo, family) were extracted from review metadata when available and used as demographic proxies. Sentiment analysis was performed using the AFINN lexicon in tidytext (Silge & Robinson, 2016) and validated with alternative methods (Nuccio and Bertacchini, 2023).

## Analytic Techniques

We employed a range of analytical methods to investigate different facets of visitor behaviour and sentiment. These included clustering algorithms, sentiment and natural language processing, graph-based spatial modelling, and visual exploratory techniques.

*Natural Language Processing (NLP)*

To analyze TripAdvisor reviews, we utilized tidytext and other R packages to tokenize, clean, and parse the text. We focused on three primary NLP tasks:

**Sentiment Analysis.** Using the AFINN and NRC lexicons, we scored each review by aggregating sentiment-bearing words. We calculated sentiment polarity (positive/negative) and strength, then compared across user types, seasons, and lag between visit and review (Nuccio and Bertacchini, 2023).

**Topic Extraction.** While not central to our analysis, we used TF-IDF (term frequency inverse document frequency) to explore review topics by user group and language.

**Language Detection and Translation.** Non-English reviews were translated and tagged for language to test for systematic bias or satisfaction variation by nationality.

*Clustering Visitor Behaviour*

To analyse the Audio Guide data, we applied unsupervised hierarchical clustering to group visitor sessions based on behaviour. Each trip was represented as a binary vector of object interactions. Pairwise distances were calculated using the Jaccard index, which measures dissimilarity in object visitation. Clustering was performed using UPGMA (Unweighted Pair Group Method with Arithmetic Mean), which was found to balance robustness and interpretability.

After evaluating dendrogram stability and silhouette width, we selected a four-cluster solution that offered clear interpretability and aligned with museum intuition. Cluster characteristics were further explored by language, time spent, and number of visited objects.

### *Spatial Flow Modelling*

We reconstructed room-to-room transitions by mapping object interactions to gallery locations from the Audio Guide data. This created a directed graph of movement probabilities between rooms. We computed transition frequencies, doorway preference matrices, and calculated path distances using random walk models. Staircases were modelled with behavioural cost penalties derived from observed avoidance rates (Hoare, 2025). We then visualised The Museum's flow structure, identifying bottlenecks, high-traffic zones, and inaccessible regions. We also used modified PageRank scores to measure relative room visibility and flow centrality.

### *Tour Usage and Object Completion*

The audio guide offered predefined thematic tours. We measured completion rates, partial engagement, and sequence deviations across the 9 different languages offered by the audio guides; English (en), French (fr), German (de), Italian (it), Japanese (ja), Korean (ko), Russian (ru), Spanish (es) and Chinese (zh). Tour paths were modelled as expected Markov chains, and completion entropy was calculated to capture how rigidly visitors followed intended orders.

### *Software and Tools*

All analyses were conducted using R and Python. Key libraries included tidytext, scikit-learn, igraph, networkx, and plotly for interactive visualisations. This multi-method approach enabled a comprehensive view of visitor engagement, combining quantitative structure with qualitative texture.

# Results

## Visitor Satisfaction Analysis

We began our results by analysing visitor satisfaction using TripAdvisor reviews, which included both numeric star ratings and textual comments. The average rating across all reviews was 4.6 out of 5, indicating generally high satisfaction. However, when disaggregated by visitor type, clear differences emerged.

As shown in Figure 1a below, business and solo travellers reported the highest levels of satisfaction, while couples consistently rated their experiences lower. Families were slightly below average, and friend groups showed wide variance. These trends remained stable across language and season.

We examined temporal effects by measuring the time lag between a user's visit and when they wrote their review. Interestingly, reviews written immediately after the visit were slightly more critical than those written weeks or months later. This suggests a recency effect, where negative logistical experiences (e.g., queues, crowds) dominate fresh memory but fade over time, leaving more lasting impressions of the cultural content.

### *Sentiment and Language Differences*

Using sentiment scoring, we found consistent trends across languages. On average, favorable terms such as beautiful, incredible, and impressive dominated English, Spanish, and French reviews alike. Some language groups, such as Japanese and German, used fewer superlatives but did not score lower in sentiment overall.

Specific keywords appeared more frequently in lower-rated reviews, including crowded, rude, queue, and confusing. In contrast, higher-rated reviews highlighted free, informative, must-see, and audio guides. These patterns confirmed that visitor mood is shaped both by infrastructure and by perceived cultural value.

### *Trip Type and Seasonality*

We grouped reviews by self-declared trip type and visit month (see Figure 1b below). Business travellers expressed higher satisfaction, possibly due to lower expectations or visits during quieter periods. Satisfaction peaked in the winter months, especially January and February, suggesting that lower crowding plays a substantial role.

### *Review Length and Detail*

There was a strong correlation between review length and positivity. Negative reviews tended to be shorter and more emotionally charged, while positive reviews were often longer and more descriptive. This asymmetry suggests that satisfied visitors are more likely to reflect at length, whereas dissatisfied ones may write in haste.

Together, these findings indicate that while overall satisfaction is high, there are meaningful differences by visitor type, season, and post-visit reflection. Addressing crowd-related complaints and enhancing infrastructure could particularly benefit segments like couples, who appear more sensitive to these issues (Ballantyne & Uzzell, 2011).

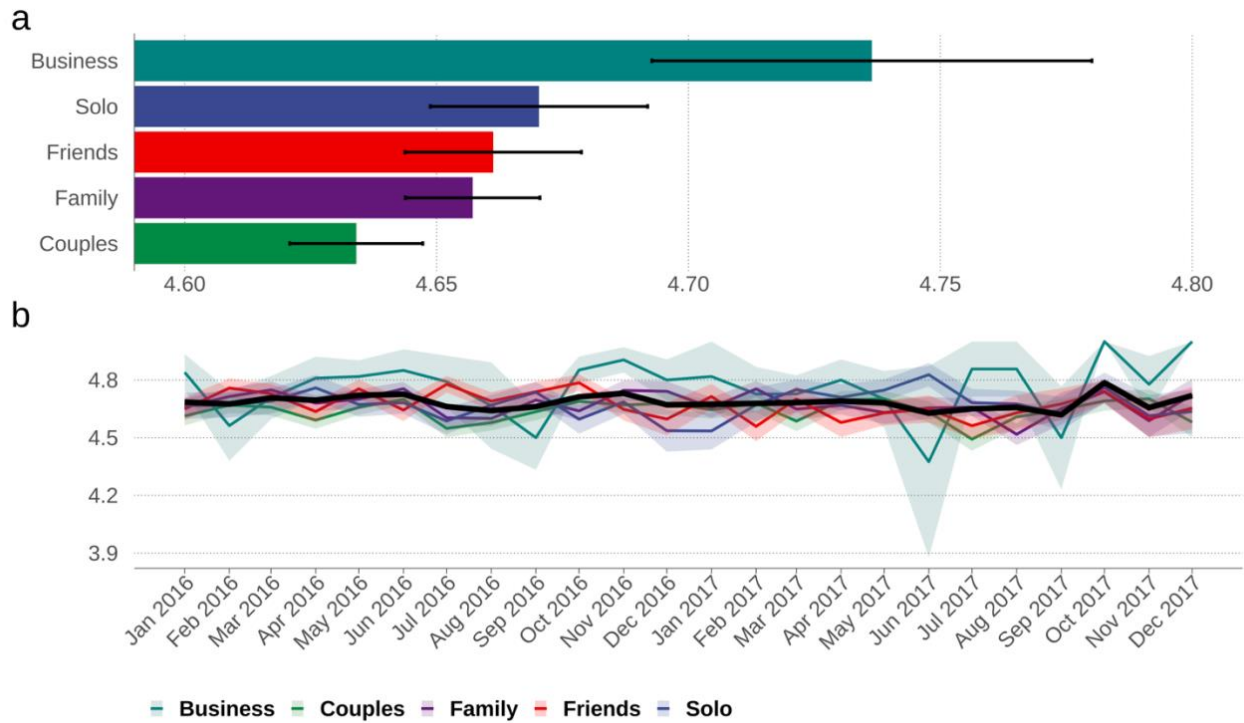

**Figure 1**: Visitor satisfaction by group type. a. Mean TripAdvisor rating by group type. Error bars denote 95% confidence intervals. b. Sentiment of visitors across time. The black line represents mean ratings across all visitors. Shading indicates 95% confidence intervals.

## Segmentation and Behavioural Clustering

While the TripAdvisor data provides insights into self-reported visitor satisfaction, attempts to analyse this data are limited both by the lack of granularity offered by TripAdvisor and traditional demographic segmentation approaches. To move beyond a reliance on demographics, we applied behavioural clustering to audio guide users to segment visitors based on their interactions with and movement through the Museum. Using a hierarchical clustering approach based on Jaccard similarity between sets of visited objects, we identified four distinct clusters of visitor behaviour. In line with the aim of creating easily digestible results to guide decision-making across the museum, we summarized the cluster into an easily understandable 'archetype', drawing inspiration from similar methodologies in geodemographic classification (see Figure 2 below). We characterised these 'archetypes' as:

**Committed Trekkers (Cluster 1).** These visitors spent significant time in the Museum, visited a large number of objects, and tended to move methodically through multiple galleries. They accounted for about 22% of sessions.

**Leisurely Explorers (Cluster 2).** These visitors had moderate-length visits, explored several themed areas, and showed signs of spontaneous browsing. They comprised around 30% of visitors.

**Targeted Visitors (Cluster 3).** This group had short visits but focused on a narrow set of iconic objects, often completing a pre-defined tour. They included many first-time and international tourists.

**Speedy Samplers (Cluster 4).** The smallest group, these users rushed through the Museum, often spending less than 20 minutes and visiting very few objects. Many appeared to activate the guide out of curiosity and quickly disengage.

Language usage varied across clusters. Committed Trekkers were disproportionately English-speaking, while Targeted Visitors and Speedy Samplers showed more diversity in languages, particularly Italian, Chinese, and French. Completion rates for tours were highest among Targeted Visitors, aligning with their focused approach.

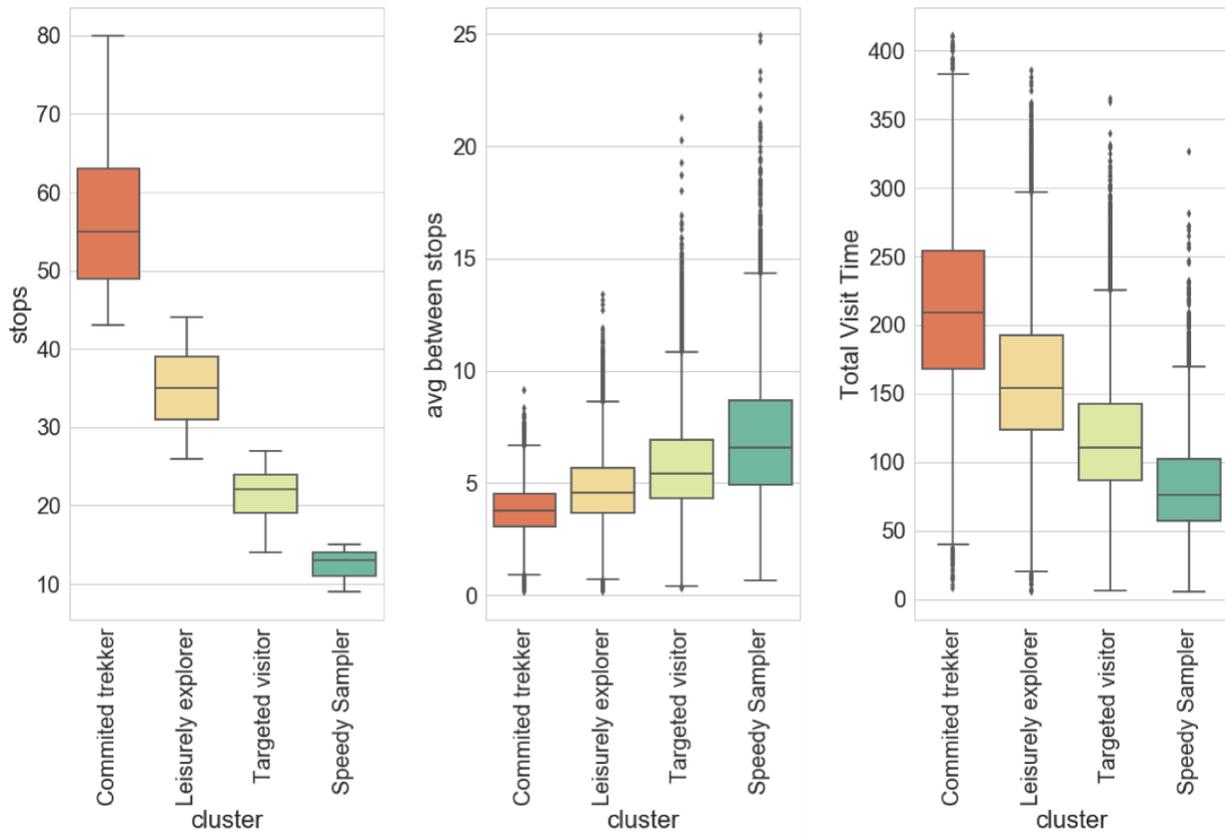

**Figure 2**: Visitor cluster comparison by objects visited, time spent, and time per object.

We compared average visit duration, number of rooms entered, and time per object. Committed Trekkers spent more than an hour on average, while Speedy Samplers typically disengaged in under 15 minutes. Leisurely Explorers averaged 30-45 minutes with moderate coverage, and Targeted Visitors showed a skewed distribution with peaks at known tour lengths.

This behavioural segmentation offers more actionable insights than simple demographic categories. For example, museum signage or mobile app recommendations can be tailored in real time: visitors who resemble Speedy Samplers could be gently moved toward slower engagement, while Targeted Visitors can be directed to complementary content they may miss. Moreover, this segmentation helps in predicting spatial density. Committed Trekkers are more likely to cause congestion in lesser-known rooms.

Such an approach aligns with a broader shift toward visitor-centred cultural strategy, as advocated in emerging museum practice. These insights point towards the utility of unsupervised clustering in cultural analytics and support a shift from one-size-fits-all assumptions toward personalised visitor services. Furthermore, such behavioural typologies can inform the design of inclusive and engaging spaces that adapt to visitor intent and stamina.

## Tour Usage and Object Interaction

The British Museum's audio guide offers several pre-set thematic tours, including Top Ten Objects, Ancient Egypt, Europe, and others in up to eleven different languages. We examined how visitors engaged with these tours, measuring stop completion rates, drop-off points, and object-level popularity across different linguistic groups (see Figure 3 below).

### *Completion Rates and Drop-offs*

We found that only 18% of visitors completed a full tour, while 52% started but abandoned it partway. Most drop-offs occurred after three or four stops, regardless of tour length. This trend suggests a mismatch between visitor expectations and tour design, especially given the physical distance between stops. Previous research has shown that wayfinding complexity and physical fatigue are major barriers to sustained cultural engagement, particularly for first-time and international visitors (Bitgood, 2006; Peponis et al., 1997).

Specific tours were particularly prone to abandonment. The Europe tour, which spans widely separated rooms, had among the lowest completion rates, along with the Korea tour. Conversely, the Top Ten Objects tour, focusing on iconic highlights in closer proximity, showed higher completion, especially among visitors using the guide in Japanese, Korean, or Chinese.

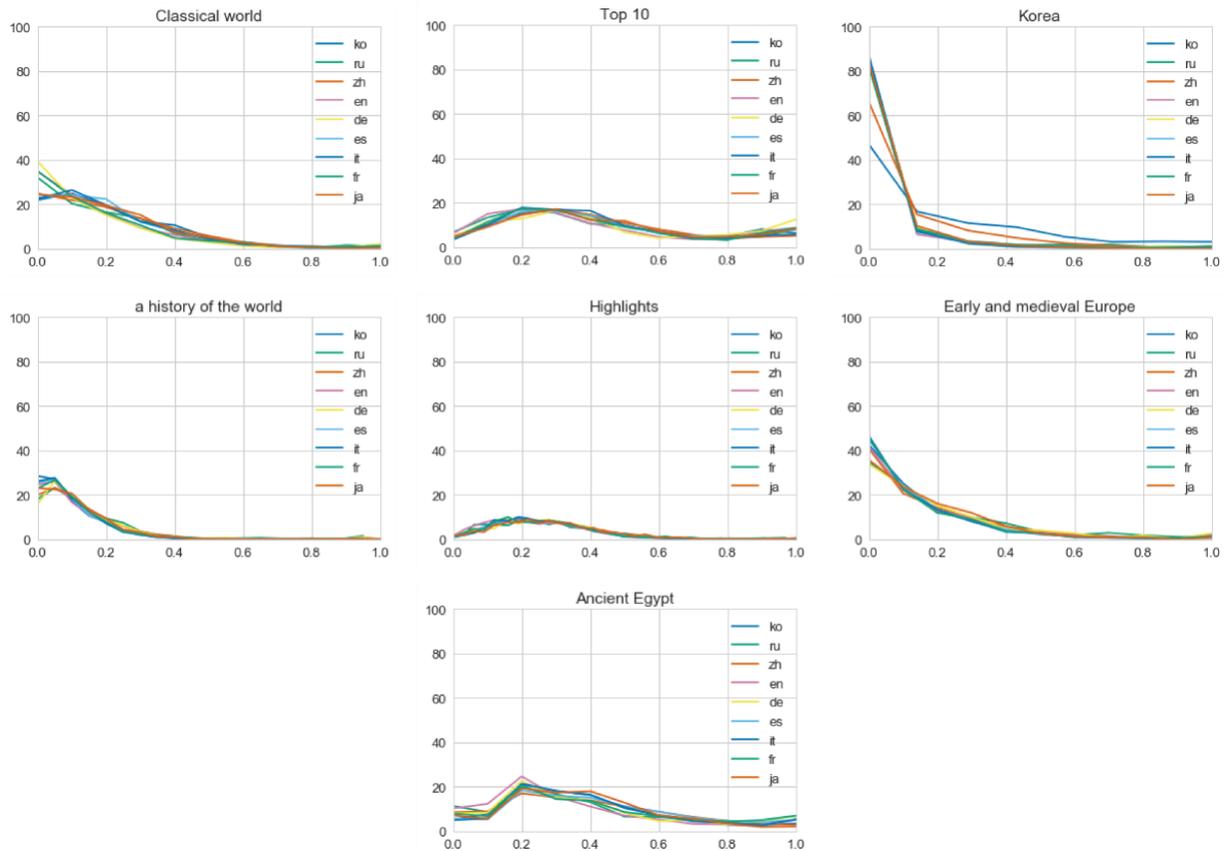

**Figure 3**: Percentage of tour stops (x-axis) completed by the percentage of visitors (y-axis) in each language group for selected tours. ko: Korean, ru: Russian, zh: Mandarin, en: English, de: German, es: Spanish, it: Italian, fr: French, ja: Japanese.

## *Language-Specific Behaviours*

Language appeared to influence both completion rates and tour preferences. English, German, and French users were more likely to browse freely after a few stops. In contrast, users of East Asian languages showed greater adherence to preset tour paths, possibly due to stronger cultural expectations around structured experiences.

The Chinese language group exhibited a unique pattern: while overall completion was low, those who did follow a tour tended to stick with it to the end. This supports targeted UX adaptations, such as offering a streamlined, minimal walking tour in Chinese or enabling visitors to set distance tolerances at tour start.

*Object Interaction*

Beyond Tours, we analysed which objects received the most engagement. Not surprisingly, the Rosetta Stone (an inscribed granodiorite stele that provided the key to deciphering Egyptian hieroglyphs), the Parthenon sculptures (classical marble reliefs from the Athenian temple depicting mythological and civic scenes), and the Egyptian mummies (preserved remains illustrating ancient beliefs about the afterlife) were among the most accessed. However, their visitation rates differed by language: while nearly all visitors saw the Rosetta Stone, the Lewis Chessmen were disproportionately accessed by English speakers, and the Sutton Hoo helmet was more popular among European language users.

*Design Implications*

These findings suggest several areas for improvement to enhance the museum experience. Tours could be redesigned to reduce the physical distance between stops, making them more accessible and less tiring for visitors. Introducing dynamic route planning within the app—for example, allowing users to select preferences like 'I have 30 minutes and want Egyptian highlights', would offer a more personalized experience. Clearly signaling tour difficulty levels, such as options that cover 10 objects in under 20 minutes with minimal walking, can help visitors make informed choices based on their energy levels and available time. Encouraging spontaneous exploration by rewarding off-tour interactions could further promote object discovery and engagement. Overall, aligning tours more closely with visitor intent, stamina, and language preferences has the potential to significantly increase engagement, reduce fatigue, and foster a deeper connection with the Museum's offerings.

## Spatial Flow and Movement Patterns

To understand how visitors navigate The Museum's physical space, we reconstructed spatial trajectories from audio guide logs. By mapping each object to its gallery room, we were able to model room-to-room transitions and generate a flow network of visitor movement (see Figure 4 below). This provided a comprehensive view of behavioural patterns over time, capturing how physical layout shapes exploration.

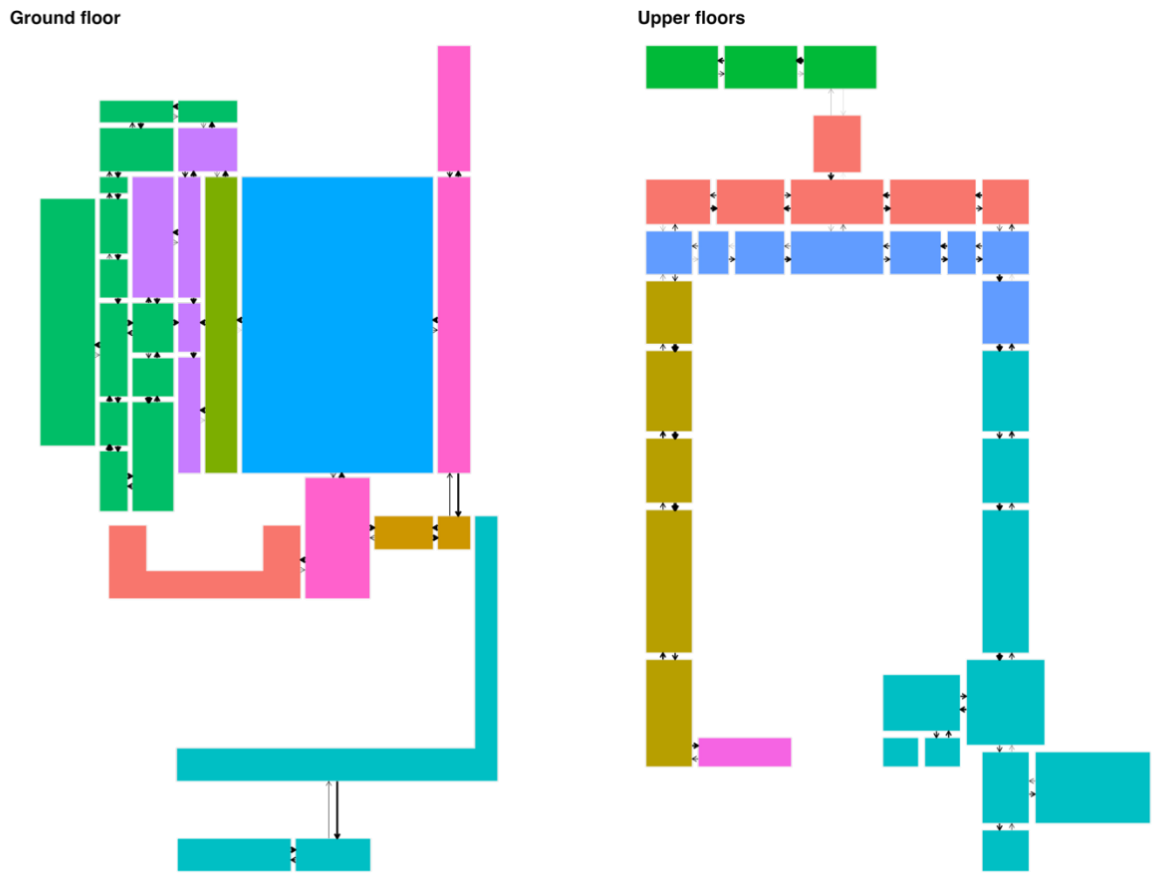

**Figure 4**: Mapping the flow of visitors across the Museum. Arrows represent frequent transitions; line thickness indicates transition volume.

**Movement Probabilities and Bottlenecks.** The vast majority of visitors began their journey in the Great Court (Room 1), where the audio guide pickup desk is located. From there, transitions were most frequent into Rooms 4 (Egyptian sculpture), 18 (Parthenon), and 24 (Rosetta Stone). These rooms formed the core of a high-traffic inner loop, visited by almost all users.

We observed bottlenecks at key junctions, particularly the west staircase, which connects ground-floor galleries to upper floors. Visitor drop-off occurred sharply after staircases, especially when ascending. The steepest drop-off occurred between Rooms 24 and 40, where nearly 70% of users who accessed Room 24 never continued upward (Falk, 2016b; Peponis and Wineman, 2010).

**Stair Aversion.** To quantify the observed aversion to stairs, we introduced a behavioural penalty in our spatial model. When we calculated the shortest paths between rooms using a random walk simulation, we added cost multipliers for stair transitions. The best-fitting model penalised upward stairs 3 more than flat transitions, while downward stairs carried little additional cost.

This adjustment improved model fit to actual flow data and suggested that vertical accessibility, not just horizontal distance, significantly deters exploration. Notably, this effect was more substantial among older visitor groups (as inferred from review language and tour patterns) and in tour-heavy languages such as Chinese and Japanese.

**Movement Patterns by Cluster.** Different visitor clusters followed different spatial patterns. Committed Trekkers were more likely to explore distant or less trafficked rooms, including Rooms 5256 (Africa and Oceania). Speedy Samplers and Targeted Visitors rarely left the high-traffic core. Leisurely Explorers showed mixed paths, often following intuitive loops that remained on one level (Bitgood, 2006).

The Museum's vertical design and room distribution inadvertently limit visitor access to several rich but under-attended galleries. Improving the visibility, accessibility, and promotion of these spaces, particularly upstairs, could distribute footfall more evenly. We suggest adding visual prompts near staircases to help visitors understand what they will see if they continue. Interactive signage could offer personalised routing: 'You've seen 5 objects in the Egyptian collection, want to discover African art upstairs in 3 minutes?' These behavioural insights provide a grounded basis for rethinking wayfinding, gallery layout, and visitor flow optimisation.

## Room Popularity Drivers

To understand what makes certain rooms more popular than others, we explored two hypotheses: that popularity is driven by a room's thematic content or by its physical location and accessibility.

We began by examining whether cultural themes (as defined by the Museum's map) influenced room popularity—Figure 5 (left below) plots room popularity by theme. While a few rooms within each theme had high visitation, the majority were less frequented, and the trend was consistent across themes. Statistical testing confirmed that the cultural theme was not a strong predictor of visitor numbers.

We then tested whether physical distance from the starting point, the audio guide collection desk in the Great Court (the central focal point of the Museum), was a better predictor. A random walk model was used to calculate room distance, producing a clear spatial gradient across the Museum.

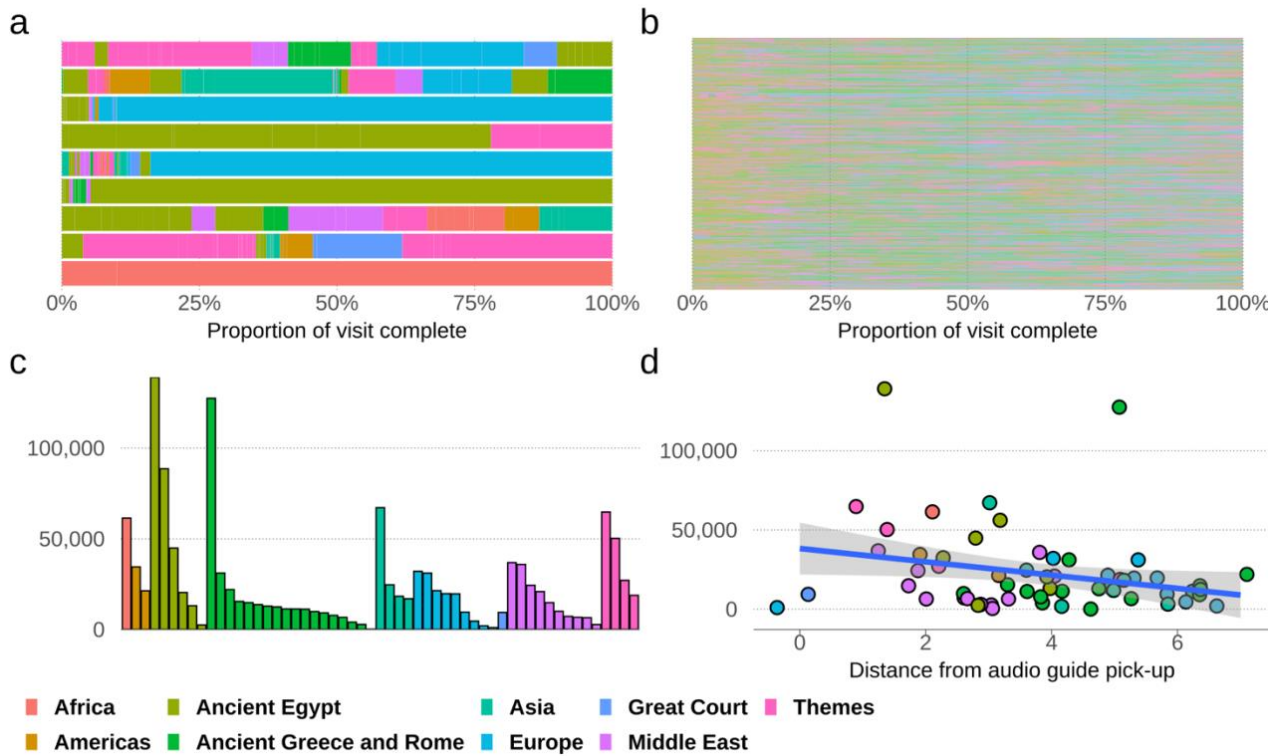

**Figure 5**: *Left*: Number of visitors to each room, coloured by cultural theme. *Right*: Room popularity as a function of distance from the starting point. Colour indicates the room theme.

When distance was plotted against visitor numbers (See Figure 5 (right above)), a significant negative correlation emerged: rooms closer to the starting point were more popular. However, there were outliers, remote rooms with high visitation and nearby rooms with low traffic, indicating other influencing factors.

To address this, we adjusted the distance metric by incorporating penalties for stair use, based on behavioral findings from earlier sections. After adjusting for the aversion to stairs, the predictive power of distance improved markedly. This confirmed that a room's physical accessibility, rather than its cultural significance, is the most consistent determinant of visitation.

Interestingly, the penalty for going upstairs was substantial, while going downstairs had minimal impact on room popularity. Visitors seem to view ascending as a disruption, even if a room is only marginally more distant.

*Design Implications*

Recommendations for boosting under-visited but valuable rooms focus on improving both physical access and visitor perception. Quieter, upstairs galleries could be signposted as tranquil alternatives, appealing to those seeking a more peaceful experience. Adding intermediate attractions or visual prompts to stairwells may encourage movement between floors, making these areas feel more connected and inviting. Revisiting room placement strategies based on predicted footfall could help distribute visitors more evenly throughout the museum. Additionally, enhancing lift signage and ensuring comfort for visitors with mobility concerns can make upper levels more accessible. This analysis suggests that optimizing physical access and reducing psychological barriers may have a greater impact on engagement than simply rebranding content or themes.

## Discussion and Conclusion

This research demonstrates the value of combining large-scale behavioural data with natural language feedback to understand and enhance visitor experiences in museums. Traditional segmentation strategies based on demographic profiles offer limited predictive power compared to clustering approaches. By analysing over 42,000 audio guide journeys and 50,000 TripAdvisor reviews, we uncovered actionable patterns in how visitors navigate, engage, and reflect on their museum experience.

Our clustering approach revealed four distinct visitor types: Committed Trekkers, Leisurely Explorers, Targeted Visitors, and Speedy Samplers, defined more by style and duration of engagement than by language or cultural background. These behavioural archetypes provide a meaningful basis for personalising tours, adjusting signage, and anticipating crowding.

Spatial movement analysis further showed that accessibility plays a critical role in shaping visitor flows. Visitors demonstrate a marked reluctance to use staircases, especially when moving upwards. Rooms located on upper floors or accessed via complex paths are significantly less visited, regardless of their thematic content (Falk, 2016a). This finding has important implications for exhibition planning and layout design.

Visitor satisfaction, as inferred from TripAdvisor reviews, is highest among business and solo travellers, and lowest among couples. Factors influencing satisfaction differ by group: solo visitors value content, while families and couples respond to amenities and logistical ease. Interestingly, reviews written long after the visit tend to be more positive, suggesting memory decay of minor inconveniences (Yalowitz and Bronnenkant, 2009).

We propose several strategies to improve the visitor experience by tailoring engagement to different needs and behaviors. Messaging should be customized for distinct visitor types—for example, promoting romantic or amenity-focused experiences to couples, while highlighting depth and variety for solo visitors and committed explorers. Spatial exploration can be encouraged by signposting upper-floor galleries as peaceful or unique destinations and enriching staircases with interpretive or visual stimuli. Tour delivery should be adapted by making audio guide tours more visible and flexible, and by using behavioral cues to suggest relevant content mid-visit. The app's interactivity can be leveraged by prompting users to declare their visit goals, such as a short browse or an in-depth exploration, and adjusting recommendations accordingly. Finally, improving perceptions of accessibility through clearer lift signage, reduced stairwell obstacles, and well-communicated effort-saving paths can help visitors feel more confident navigating the space.

Our findings reinforce the potential value of data science in the cultural sector. Museums are rich, complex environments where the visitor experience hinges on both content and context. By understanding not just who visitors are but how they behave, institutions can more effectively design inclusive, engaging, and efficient spaces. The methods demonstrated here, ranging from NLP to spatial analytics, are scalable to other museums and heritage institutions, offering a framework for evidence-based decision-making in cultural settings (Kelly, 2010; Ridge, 2014).

Future work may explore real-time personalisation, integration of Wi-Fi or Bluetooth tracking data, or expanded A/B testing of signage and layouts. We also encourage further efforts to combine behavioural data with qualitative research to deepen our understanding of motivation, emotion, and memory in cultural engagement.

## Acknowledgment

We thank Coline Cuau, Joanna Hammond, Harrison Pim, Natalia Hudelson, Hannah Boulton, and Emily Carmichael from the British Museum team for facilitating this research and their insightful comments. We thank the Alan Turing Institute for financial support.

BEHAVIOURAL SEGMENTATION AT THE BRITISH MUSEUM                               19